\documentclass[aps,12pt,longbibliography]{revtex4-1}
\usepackage[lofdepth,lotdepth]{subfig}
\usepackage{graphicx}
\usepackage{color}

\newcommand{\degree}{^{\rm o}}

\begin{document}
%

\title{Modeling of Airgap Influence on DC Voltage Generation in a Dynamo-Type Flux Pump}
\markboth{Modeling of Airgap Influence on DC Voltage ...}{}

\author{Asef Ghabeli, Enric Pardo\footnote{Author to whom correspondence should be addressed.\\Email addresses: asef.ghabeli@savba.sk, enric.pardo@savba.sk.}}

\address{Institute of Electrical Engineering, Slovak Academy of Sciences, Dubravska 9, 84104 Bratislava, Slovakia}


\date{\today}

\begin{abstract}

High-temperature superconducting (HTS) Flux pumps are promising devices to maintain steady current mode in HTS magnets or to energize rotor windings in motors and generators in a contactless way. Among different types of flux pumps, the dynamo-type flux pump has been very common due to its simple structure and ease of maintenance. However, understanding the principle of dynamo-type flux pump has been challenging despite the recent progress. For addressing this challenge, the numerical modeling has proved to be an appropriate tool. Modeling is usually fast, precise, cost-efficient and enables to examine some details that is hardly possible via measurements. For this purpose, an efficient numerical model based on Minimum Electromagnetic Entropy Production (MEMEP) method  has been used to study the performance of the flux pump in open-circuit mode. This model is the fastest and most efficient model of a flux pump with good agreement with experiments. In addition to the main behavior of the flux pump, the influence of airgap on open-circuit DC voltage of the flux pump has been investigated. This study revealed an important property in HTS flux pumps that with increasing airgap, pumping voltage in superconducting tape does not cease, but only reaches to insignificantly low values, which is not measurable via experiments. Furthermore, the modeling results of open-circuit DC voltage were compared to experimental ones obtained from the article published in 2016 by Bumby \textit{et al.} \cite{bumby2016anomalous} and the results showed good agreement. 

\end{abstract}

\maketitle
\section{Introduction}

Recent progress in the manufacturing technology of high-temperature superconducting (HTS) coated conductors (CC), paved the way for utilizing HTS CCs in magnets and power electric devices such as motors, generators and fault current limiters. The new generations of HTS conductors, HTS 2G conductors, are possible to be utilized under high critical current density ($ J_{c} $) and magnetic field ($ B $). However, applications with windings have been impeded in part because of thermal losses at the current leads and joule loss at joints. Although there has been some progress in the superconducting joint technology \cite{patel2015mgb2},\cite{park2014superconducting}, \cite{matsumoto2017superconducting}, there is still a long way to develop reliable and consistent superconducting joints with the similar HTS CC characteristics such as critical current, resistance and magnetic field dependence of critical current. In synchronous motors and generators, powering the pole coils at the rotor requires slip rings or brushes, adding additional problems to the aforementioned thermal loss. In addition, thick current leads in HTS motors and generators also imposes a considerable amount of heat and joule loss. 

HTS flux pumps are devices that can be considered to address these problems. They can be utilized to energize the HTS magnets to maintain the steady current mode or in rotor windings of electrical motors and generators to inject the current in a contactless way without using brushes. Another reason to avoid using brushes, which are a crucial part of off-shore wind turbines, is that they require continuous maintenance, which imposes extra costs.    

In general, there are three main types of HTS flux pumps including traveling wave flux pumps, transformer-rectifier and variable resistance rectifier flux pumps \cite{coombs2016overview}. In recent years, the travelling wave flux pumps have been very popular because of their simplicity, not using any switches and hence less maintenance and cost. Among the travelling magnetic wave flux pumps, the rotating magnetic wave flux pumps that use magnets for creating varying magnetic field are even more simple and easy to maintain. These kinds of flux pumps are called dynamo-type flux pumps. Unlike the linear electro-magnet based travelling wave flux pumps, which utilize power electronic devices to generate varying magnetic field, the rotating ones do not require such equipment and use magnets to create the magnetic field. 

	The principle of a travelling magnetic wave flux pump can be simply illustrated as follows: a varying magnetic field relative to the HTS tape starts to penetrate inside the HTS tape. This penetrated magnetic field couples with the clusters of vortices inside the tape and with the movement of magnetic poles the coupled vortices will start to move accordingly, which eventually leads to loss generation inside the superconductor \cite{wang2018macroscopic}. This interaction occurs in the flux flow regime. It means that unlike the previous generations of flux pump that were employing low-temperature superconductors and also $ \rm{MgB_{2}} $ \cite{van1981fully} there is no need to drive the superconductor to the normal regime. As seen in \cite{mataira2019origin}, the only necessary condition to have the flux pumping or non-zero DC voltage component is having non-linear resistivity in the material, and hence the presence of vortices is not a requirement.

 	Since 2011 when Hoffmann and \textit{et al.} proposed and designed the first travelling wave flux pump based on rotating magnet (dynamo-type HTS flux pump) \cite{hoffmann2010flux}, \cite{walsh2013characterization}, \cite{hoffmann2012design}, several articles have been published on performance, design and optimization of this type of flux pump. In \cite{jiang2015impact}, it was demonstrated that the function of HTS dynamo can be described by a simple circuit model including open-circuit voltage and internal resistance of the flux pump, which are closely proportional to frequency. In addition, the flux gap between magnets and HTS tape is a critical factor. In \cite{bumby2016anomalous}, the performance of a simple HTS dynamo in open-circuit condition was examined  and explained with a previously proposed model by Giaever \cite{giaever1965magnetic}. An innovative method was employed in \cite{jiang2016novel} to concentrate all the rotor magnets flux on the CC wire by designing a ferromagnetic circuit comprised of teeth on the stator yoke. Using this method, a prototype of dynamo-type flux pump entirely outside the cryogenic system was successfully demonstrated \cite{bumby2016through}. The impact of stator wire width in an HTS dynamo type flux pump has been investigated in \cite{pantoja2016impact}. It is shown that the ratio of magnet width to stator wire width is a key factor for its design. In \cite{bumby2016development}, a flux pump exciter prototype has been implemented upon a 10 kW HTS synchronous generator and tested successfully. Badcock and \textit{et al.} have shown that the efficiency of an HTS dynamo increases with a total magnet area and does not depend on the magnet orientation \cite{badcock2016impact}. Hamilton and his coworkers reported on the design and successful testing of an high-efficiency squirrel-cage HTS dynamo with the output of 700 A, which can be raised up to 1.3 kA. They also showed that the highest current value would achieve if the number of rotor magnet is less than half of the stator wire number \cite{hamilton2018design}. The Effect of frequency on output DC voltage of an HTS dynamo has been studied in \cite{pantoja2018output}. It was found that there are three frequency regimes of behavior including low, mid and high regions. Only in low frequency region, the output DC voltage is directly proportional to frequency, while in mid and high frequency regimes, the output DC voltage is roughly constant and drops with increasing the frequency, respectively. In \cite{storey2019optimizing}, the effect of the number of rotor magnets and rotor speed on the performance of an HTS dynamo was examined. A linear relation was observed with the number of magnets and output $V_{oc}$ and also short-circuit current. However, an unexpected sublinear dependence was noticed with magnet speed, which was explained as the effect of eddy current flowing on the stator yoke. Hamilton and coworkers in 2019 studied the effect of synchronous and asynchronous number of magnets on the rotor of an HTS dynamo\cite{hamilton2019asynchronous}. They observed that both cases lead to efficient output voltage and current, but they recommended the asynchronous configuration because it generates more constant output current. 

	There have been few papers that modeled the performance of HTS flux pump. In 2017, a simple yet effective linear HTS flux pump was modeled with finite element calculations using a 2D $A$-formulation \cite{campbell2017finite}. The model could successfully demonstrate the pumping performance of the flux pump although there were some drawbacks. The first one was using critical state model with constant $ J_c $ for defining superconductivity, so it lacked consideration of $ J_{c}(B) $ dependency, which is crucial in the process of flux pumping. The other was not using a series resistance or inductance load connected to the superconducting tape, which led to very fast saturation of load in few cycles. In addition, using very limited time steps and also sudden vanishing of magnet to generate pumping behavior made the model unrealistic and far from experimental flux pump. In \cite{li2017numerical} and \cite{ainslie2018numerical}, the dynamic resistance in HTS coated conductors have been modeled using T-formulation and H-formulation in 2D, respectively. Both models were validated using experimental measurements. However, since they have used stationary sinusoidal ac magnetic field as external applied magnetic field, the case is totally different with external magnetic field due to the moving magnet, so these models cannot provide a realistic model of the mechanism of a dynamo-type flux pump. In \cite{geng2018modeling}, a transformer-rectifier HTS flux pump was modeled in COMSOL using a 2D H-formulation and verified by experimental data. It was demonstrated how dynamic resistance is created using traveling magnetic field. In 2019, Mataira and \textit{et al.} proposed the first model of a dynamo-type HTS flux pump using a 2D H-formulation finite element method in COMSOL \cite{mataira2019origin}. They calculated the DC output voltage in open-circuit condition and showed that this DC voltage is due to the generated eddy currents in over-critical condition, which forms the dynamic resistance. They used \textbf{E}-\textbf{J} power law and $ J_{c}(B,\theta) $ dependency of critical current density for accurate calculation of output voltage and better matching with experimental data. 

	Most of the Finite Element Methods (FEM) used for simulating superconductors including tapes, coils, and stacks (such as methods utilizing commercial software) employ boundary conditions at the external surface of the air domain. The air domain should be large enough (around 10 times bigger than the dimension of superconducting object) in order to obtain a precise solution, and hence it must be meshed \cite{ghabeli2015optimization}, \cite{ghabeli2015novel}. This adds a significant number of mesh elements to the solution and thus significant extra computing time. Additionally, in the case of simulating a moving object (in this case a moving magnet), the moving mesh should be employed in the air domain, which imposes a lot more complexity and computation time to the problem. However, using minimum electromagnetic entropy production method (MEMEP), which has been used for modeling in this paper, the problem and the meshed object have been limited to only the superconducting part, hence, there will be a significant reduction in computation time and complexity.

	In summary, due to the complexity of performance of travelling magnetic wave flux pump and in particular dynamo-type HTS flux pump, the details of its principle have not been fully understood yet \cite{pardo2017dynamic}.  Moreover, an efficient, fast and yet precise modeling method seems to be a reasonable choice due to the following reasons: restrictions in experimental studies like measurement errors, difficulty in measuring some parameters such as magnetic field density and voltage, when the signal getting too weak to be measured, and most importantly the cost of designing and building apparatus along with conducting the studies. Besides, for better analyzing the complex principle of flux pump, studying the simpler cases such as open-circuit mode can be more suitable due to eliminating the effect of the current flowing in the HTS circuit. 
	
	For this purpose, in the present paper, a modeling method based on MEMEP was employed to demonstrate the performance of a dynamo-type flux pump and systematic study of the gap dependence. For verification of the modeling results, they were compared with the experimental results obtained from \cite{bumby2016anomalous}. The comparison shows good agreement with measurements, which emphasizes the efficiency of the presented modeling.


\section{Modeling Methodology}
\subsection{MEMEP 2D method}

In this paper, the MEMEP method in two dimensions has been employed for modeling. This variational method can be specially used for materials with nonlinear $ \textbf{E}(\textbf{J}) $ relation such as superconductors. It performs based on the calculation of current density $ \textbf{J} $ and scalar potential $ \varphi $ by minimizing a functional containing all the variables of the problem such as magnetic vector potential $\textbf{A}$, current density $\textbf{J}$, and scalar potential $\varphi$. It has been proved that the minimum of this functional in the quasimagnetostatic limit is the unique solution of Maxwell differential equations \cite{pardo20173d}. Since the current density is only inside the superconducting part, discretization of mesh in the domain around the superconducting part is not needed in this method. The general equation for the current density and the scalar potential are as follows: 

\begin{equation}
\textbf{E}(\textbf{J})=-\dot{\textbf{A}}-\nabla\varphi
\label{eq1}
\end{equation}  

\begin{equation}
\nabla\cdot\textbf{J}=0
\label{eq2}
\end{equation}  

The vector potential $\textbf{A}$ in the equation has two contributions including $\textbf{A}_{a}$ and $\textbf{A}_{J}$, which stand for vector potential due to applied field and vector potential due to the current density in superconductor, receptively. In here, 2D geometry has been assumed so that $ \textbf{J}=J\textbf{e}_{z} $ and $ \textbf{A}=A{\textbf{e}_{z}} $. Furthermore, the coulomb's gauge ($ \nabla\cdot \textbf{A}=0 $) has been assumed in infinitely long problems for which \cite{6648727}:

\begin{equation}
\textbf{A}_{J}(\textbf{r})=-\frac{{\mu}_{0}}{2\pi} \int_{S}d S' J(\textbf{r}')\:\ln|\textbf{r}-\textbf{r}'|
\label{eq2.1}
\end{equation}

For cross-sectional 2D problems with current constraints, equation (\ref{eq2}) is always satisfied, so only equation (\ref{eq1}) needs to be solved. For solving this equation, the following functional needs to be minimized \cite{pardo20173d}, \cite{pardo2015electromagnetic}: 

\begin{equation}
L=\int_{S}ds[\frac{1}{2} \frac{\Delta \textbf{A}_{J}}{\Delta t}\Delta \textbf{J}+ \frac{\Delta \textbf{A}_{a}}{\Delta t}\Delta \textbf{J}+U(\textbf{J}_{0}+\Delta \textbf{J})]
\label{eq3}
\end{equation}
where $U$ is dissipation factor defined as \cite{pardo20173d}:

\begin{equation}
U(\textbf{J})=\int_{0}^{J}\textbf{E}(\textbf{J}').d\textbf{J}'
\label{eq4}
\end{equation}

This dissipation factor can include any $\textbf{E}-\textbf{J}$ relations in superconductors including the critical state model \cite{pardo20173d}.

	The functional is minimized in discrete time steps. If the functional variables are $\textbf{J}_{0}$, $ \textbf{A}_{J0} $ and $\textbf{A}_{a0}$ in a particular time step like $t_{0}$, then on its next time step $t=t_{0}+\Delta t$, the variables will become $\textbf{J}=\textbf{J}_{0}+\Delta \textbf{J}$, $\textbf{A}_{J}=\textbf{A}_{J_0}+\Delta \textbf{A}_{J}$ and $\textbf{A}_{a}=\textbf{A}_{a0}+\Delta \textbf{A}_{a}$, respectively, where $\Delta \textbf{J}$, $\Delta \textbf{A}_{J}$ and $\Delta \textbf{A}_{a}$ are the change of the variables between two considered time steps and $\Delta t$ is the time period between the two time steps. In this modeling, the non-uniform applied magnetic field $\textbf{B}_{a}$ caused by the rotating magnet appears in the functional in the form of $\textbf{A}_{a}$ \cite{pardo20173d}. 
	As previously explained, meshing was performed only inside the superconductor region in rectangular shape elements and uniformly across the object. For this modeling, only one element along the thickness of the superconducting tape suffices and the optimum number of elements along the width was chosen to be around 200 elements. This number has been increased up to 1000 elements, when higher precision is needed, especially in larger airgaps where the magnetic field density is weak. Therefore, the whole number of elements in the modeling was between 200 to 1000 elements, which reduces the computation time significantly compared to conventional FEM. Besides, the optimum number of time steps that were chosen were 360 per cycle. For this number of mesh and time step, the computation time was less than 20 minutes per cycle.  

\subsection{Magnet Modeling}

The vector potential and magnetic flux density generated by the magnet can be calculated by the magnetization sheet current density $ \textbf{K}=\textbf{M} \times \textbf{e}_n $, where $ \textbf{M} $ is the magnet magnetization and $ \textbf{e}_n $ is the unit normal vector to the surface. For uniform magnetization, the vector potential $ \textbf{A}_M $ and magnetic flux density $ \textbf{B}_M $ generated by the magnet are: 

\begin{equation}
\textbf{A}(\textbf{r})=-\frac{\mu_0}{2\pi} \, M \int_{ \partial S} dl' \, \textbf{e}_m \times \textbf{e}_n(\textbf{r}') \ln |\textbf{r}-\textbf{r}'|
\end{equation}

\begin{equation}
\textbf{B}(\textbf{r})=\frac{\mu_0}{2\pi} \, M \int_{ \partial S} dl' \, \frac{[\textbf{e}_m \times \textbf{e}_n(\textbf{r}')]\times (\textbf{r}-\textbf{r}')}{|\textbf{r}-\textbf{r}'|^2}  
\end{equation}
where $ \textbf{e}_m $ is the unit vector in the direction of magnetization, $ \partial S $ represents the edges of the magnet cross-section, and $dl'$ is the length differential on the edge. Note that since both $ \textbf{e}_m $ and $ \textbf{e}_n $ are in the $ xy $ plane, the cross-product $ \textbf{e}_m \times \textbf{e}_n $ is always in the $z$ direction, and hence $\textbf{A}$ follows the $z$ direction.

In this article, the $\textbf{A}_M$ and $\textbf{B}_M$ were evaluated numerically. 

The magnetic field generated by the magnet acts as an applied field. Then, the program only needs to calculate this magnetic field, once for each time step within the first cycle. The impact on the total computing time is negligible, since minimization takes most of the computing time.

\subsection{HTS Tape Characteristics}

	For defining the non-linear superconductor characteristic, the model uses the isotropic $\textbf{E}-\textbf{J}$ power law:
	
\begin{equation}
\textbf{E}(\textbf{J})=E_{c}\left(\frac{|\textbf{J}|}{J_{c}}\right)^n \frac{\textbf{J}}{|\textbf{J}|}
\label{eq5}
\end{equation}                                 

where $E_{c} =10^{-4} \; V/m$ is the critical electric field , $J_{c}$ is the critical current density and $n$ is the power law exponent or n-value. 

With this $ E(J) $ relation, the non-linear resistivity is $ E(J)/J $ and hence:

\begin{equation}
\rho(\textbf{J})=\frac{E_{c}}{J_{c}} \left(\frac{|\textbf{J}|}{J_{c}}\right)^{n-1}
\label{eq5.1}
\end{equation}

For the $\textbf{E}-\textbf{J}$ power law, the disspiation factor of the functional in the Equation (\ref{eq4}) becomes:

  \begin{equation}
U(\textbf{J})=\frac{E_{c}J_{c}}{n+1} \left(\frac{|\textbf{J}|}{J_{c}}\right)^{n+1} 
\label{eq6}
\end{equation}                          

	The $J_{c}(\textbf{B})$ dependence of the critical current of HTS tape has been obtained from measurements in the temperature of 77.5$\degree$ Kelvin (liquid nitrogen) and has been utilized in the model \cite{mataira2019origin}. The critical current was measured in the range between 0$\degree$ to 180$\degree$ and the critical current dependence was assumed symmetrical from -180$\degree$  to 0$\degree$ .   

\subsection{Model Geometry}
The current density $\textbf{J}$ flows in the z direction, and hence the vector potential $\textbf{A}$ and electric field $\textbf{E}$ are parallel to $\textbf{J}$. The magnetic field $\textbf{B}$ is perpendicular to the current density and its direction belongs to the xy-plane. Due to infinitely long symmetry the following relations can be derived:

\begin{eqnarray}
\textbf{E}(\textbf{r})=E(x,y)\textbf{e}_{z}\\
\textbf{J}(\textbf{r})=E(x,y)\textbf{e}_{z} \nonumber \\
\textbf{A}(\textbf{r})=A(x,y)\textbf{e}_{z} \nonumber
\label{eq6.1}
\end{eqnarray}

\begin{figure}[tbp]
\centering
{\includegraphics[trim=0 0 0 0,clip,width=8 cm]{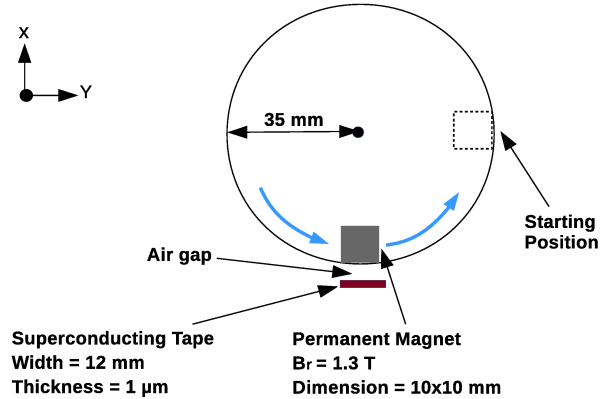}}
\caption {Configuration of 2D model} 
\label{Fig. 1}
\end{figure}

Taking these relations into account and Equation (\ref{eq1}), it is finded that:

\begin{equation}
\nabla\varphi(\textbf{r})=\partial_{z}\varphi \textbf{e}_{z}
\label{eq7}
\end{equation}                                
where $\partial_{z}\varphi$ is uniform within the superconductor. Since coulomb's gauge is assumed, the scalar potential becomes the electrostatic potential.  
Fig. \ref {Fig. 1}, shows the sketch of the modeling configuration. 

\subsection{About the Open-Circuit Voltage} \label{About the Open-Circuit Voltage}
For calculating open-circuit voltage, the gradient of the electrostatic potential needs to be calculated from Equation (\ref{eq7}).                             

In 2D modeling, the fact that the gradient of the electrostatic vector potential is uniform along the z axis, allows us to easily calculate the gradient per unit length. Afterwards, the voltage (or open-circuit voltage in our case) can be simply calculated by:

\begin{equation}
V = -\left(\frac{\partial \varphi}{\partial z}\right). \:l
\label{eq9}
\end{equation}                                
where $l$ is the length of the tape in between the voltage taps, which in the presented 2D modeling is related to the dimension of the used magnet. 

\section{General Behaviour of Open-Circuit Voltage}
This section presents results of the open-circuit voltage calculation. The parameters used for these calculations were obtained from \cite{bumby2016anomalous} so that it can be comparable to the experimental results presented in the mentioned paper. 

\subsection{Parameter Description \label{section 3.1}}
The permanent magnet used for this study is a square magnet with the dimension of 10$\times$10 mm. Note that although in \cite{bumby2016anomalous} a cylindrical shape of magnet has been used,  the conversion of a cylindrical shape in 2D would be a rectangular shape, which of course is not a perfect transformation, but accurate enough for our estimations. Furthermore, as shown in \cite{mataira2019origin} the flux pumping mechanism can be well explained by a 2D model. The magnet type is a N42 magnet, which offers the remanence magnetic field of around 1.3 T. 

	The parameters that have been selected for modeling the YBCO superconducting tape are 12 mm width and 1 $\mu$m thickness, n-value=30 and critical current at self field is 281 A. The $Jc(B,\theta)$ data used for HTS tape were obtained from the experimental data in \cite{mataira2019origin}, which belongs to the same type of the HTS tape (not identical tape) as in \cite{bumby2016anomalous}.
	
	The rotating radius of the magnet rotor is 35 mm and the magnet has been placed on the circumference of this rotor. The rotating frequency is 12.3 Hz. The set-up configuration has been shown in Fig. \ref {Fig. 1}.
	
\subsection{Main Behavior}	

The results presented in this section were calculated for airgap of 3.3 mm. The airgap was defined as the distance between the magnet surface and the HTS tape. Using the procedure explained in section \ref{About the Open-Circuit Voltage} the voltage was calculated considering the $J_{c}(B)$ dependence. Besides, all the results belong to the second cycle to skip the transient state in the first cycle. We tested and verified that the results at the following cycles are the same as those at the second cycle. 

%

\begin{figure}[tbp]
\centering
{\includegraphics[trim=0 0 0 0,clip,width=8 cm]{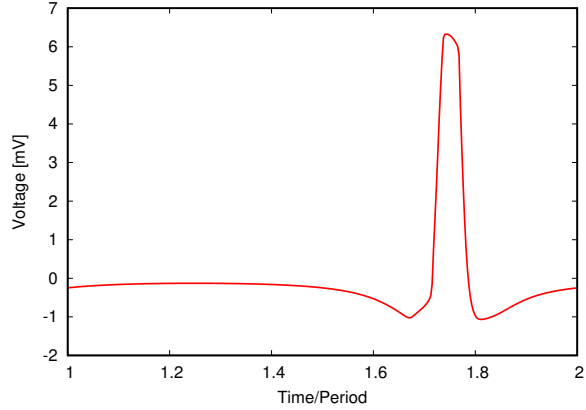}}
\caption {Output open-circuit voltage in the airgap of 3.3 mm } 
\label{Fig. 2}
\end{figure}

The calculated output $V$ is comprised of two terms: 

\begin{equation}
V=-l\:\partial_{z}\varphi=l\:\partial_{t}A+l\:E(J)
\label{eq11}
\end{equation}                                

where $A$ is the total magnetic vector potential and $E$ is the electric field, which is obtained by the $\textbf{E}-\textbf{J}$ relation considered for the HTS tape and is dependent on the local current density of the tape. For the modeled flux pump, this voltage $V$ has been calculated in Fig. \ref {Fig. 2} during the second cycle.

However, an important parameter for flux pumping is the dc voltage component, $V_{dc}$. This dc component is as follows:

\begin{equation}
\label{eq12}
V_{dc}=fl \int_{0}^{1/f} [\partial_{t} A + E(J)]dt  
=fl \int_{0}^{1/f} E(J)dt
\end{equation}                                      
where $l$ is the tape length and $\rho$ is the non-linear resistivity of HTS tape. In the equation above, the vector potential in steady state is used, which is periodic and hence $ \int_{0}^{1/f} \partial_{t}A=A[1/f]-A[0]=0 $. The reason for the periodicity of vector potential is that both $A$ from the magnet and the currents in the superconductor tape are periodic after the first cycle. 
 
\begin{figure}[tbp]
\centering
{\includegraphics[trim=0 0 0 0,clip,width=8 cm]{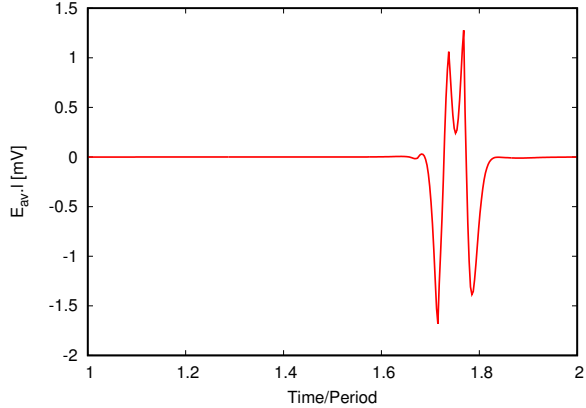}}
\caption {The generated open-circuit voltage by the non-linear resistivity of the HTS tape in the airgap of 3.3 mm. $E_{av}$ is the cross-section average of $ E(J)=\rho(J)J $ and $l$ is the tape length. } 
\label{Fig. 3}
\end{figure}

Based on Equation (\ref{eq12}), the output $V_{dc}$ only depends on the electric field generated by the resistivity of HTS tape, which itself is a function of the tape current density. An interesting feature is that $V_{dc}$ can be calculated from the time integral of $\rho(J)J$ on any point and the integrals for all points yielding the same result. Then, we can also use the cross-section average electric field to calculate $V_{dc}$, being $ E_{av}(t)= 1/S \int_{S} ds\, \rho \big(J(\textbf{r})\big) J(\textbf{r}) $ where $S$ is the cross-section surface. For the modeled flux pump, this contribution of voltage in HTS tape is shown in Fig. \ref {Fig. 3}.

The output voltage of a flux pump can be expressed as:
\begin{eqnarray}
\label{eqVdc}
V(t)=l[\partial_t A+ E (J)]=l\frac{1}{S}\int_{S}dS[\partial_t A+E(J)] =l[\partial_t A_{av}+E_{av}(J)] 
\end{eqnarray}
where $A_{av}$ is the average total vector potential in the tape and $A=A_a+A_J$ where $A_a$ is the vector potential due to external magnetic field and $A_J$ is the vector potential due to local current density in the tape. Using the same arguments as Equation (\ref{eqVdc}), we find that the dc voltage follows:

\begin{equation}
\label{VdcEav}
V_{dc}=fl \int_{0}^{1/f}E_{av}(J)
\end{equation}

 For a linear material we have:
 
\begin{equation}
E_{av}(J)=\rho J_{av}=\rho \frac{I}{S}
\end{equation}

And hence $V_{dc}$ vanishes for the open-circuit case because $ I=0 $.

The output voltage difference between the tape at 77$\degree$ K (superconductivity mode) and at 300$\degree$ K (normal mode) is:
\begin{equation}
\label{DeltaV}
\Delta V=V_{77 \degree K}-V_{300 \degree K}
\end{equation}
which is equal to
\begin{eqnarray}
\Delta V=l\left[\partial_t(A_{av,J,77\degree K}-A_{av,J,300\degree K})+E_{av}(J)- \rho_{300\degree K} \frac{I}{S}\right]
\end{eqnarray}

For open-circuit mode, the last term vanishes. In addition, since the metal resistivities are large at 300$\degree$K, the vector potential generated by currents at 300$\degree$K ($A_{av,J,300\degree K}$) will be negligible compared to those from superconductor at 77$\degree$K ($A_{av,J,77\degree K}$). Therefore we will have:
\begin{equation}
\label{deltav}
\Delta V_{oc} \approx l[\partial_t A_{av,J}+E_{av}(J)]
\end{equation}
where the sub-index $oc$ denotes the open-circuit mode. 

\begin{figure}[tbp]
\centering
{\includegraphics[trim=0 0 0 0,clip,width=8 cm]{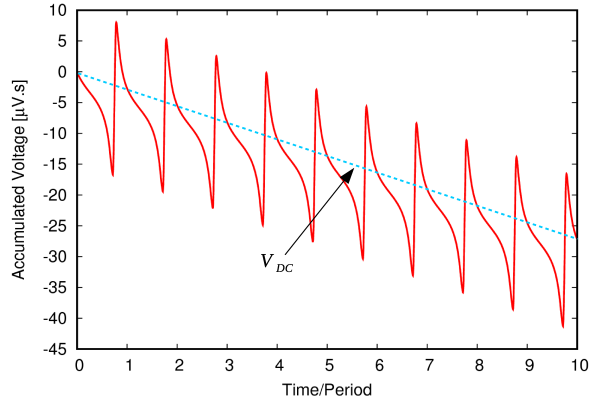}}
\caption {Accumulated open-circuit voltage over 10 cycles in the airgap of 3.3 mm } 
\label{Fig. 4}
\end{figure}

If $V_{dc}$ of the graph in Fig. \ref {Fig. 3} is calculated using Equation (\ref{VdcEav}), it can be observed that the flux pump has a DC value equal to 33.6 $\mu$V, which is responsible for energizing the flux pump. Over many cycles, this value of DC voltage will be accumulated to inject the magnetic flux into the superconducting circuit \cite{bumby2016anomalous}. In Fig. \ref {Fig. 4}, this trend can be noticed over 10 cycles for the modeled flux pump. The accumulated voltage is calculated by:

\begin{equation}
V_{accumulated}(t)= \int_{0}^{t}V_{dc}(t') \:dt' 
\label{eq13}
\end{equation} 
  
\subsection{Comparison of the Cases with Constant  $J_{c}$ and $J_{c}(B,\theta)$ Dependent}

\begin{figure}[tbp]
\centering
{\includegraphics[trim=0 0 0 0,clip,width=8 cm]{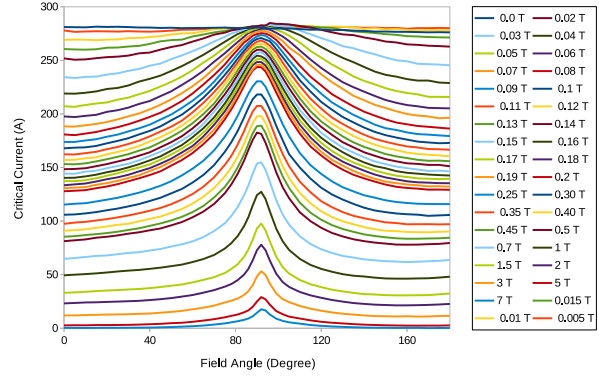}}
\caption { Experimental $I_{c}(B,\theta)$ data used in the modeling. Data was measured at 77.5 $\degree$K in magnetic fields up to 7 T, derived from \cite{mataira2019origin} } 
\label{Fig. 5}
\end{figure}

\begin{figure}[tbp]
\centering
{\includegraphics[trim=0 0 0 0,clip,width=8 cm]{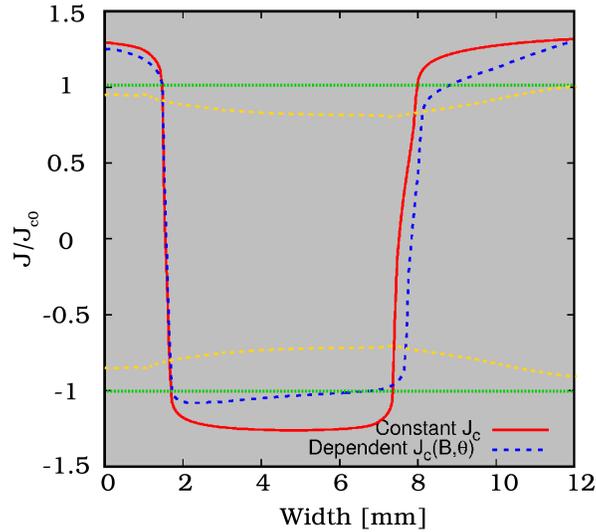}}
\caption {Comparison of the current density line diagram of constant $J_c$ (red solid line) and dependent $J_{c}(B,\theta)$ (blue dashed line) when the magnet is just on the top of the tape with the airgap of $3.3$ mm. The green dashed lines show critical current density level during assumption of constant $J_{c}$ case and the yellow dashed lines show critical current density level during assumption of $J_{c}(B,\theta)$ dependent case} 
\label{Fig. 6}
\end{figure}

The performance of the flux pump is also studied without considering $J_{c}(B,\theta)$ dependence and hence with the assumption of constant $J_{c}$ at self-field. For the latter, the generated $V_{dc}$ in each cycle would be almost half of the previous case and equal to 16.2 $\mu$V. The reason for this significant difference can be explained by the fact that the flux pumping phenomena occurs because of overcritical eddy current generated in HTS tape during traversing magnet over the tape \cite{mataira2019origin}. In addition, for instance, in the case of airgap equal to 3.3 mm, the maximum perpendicular magnetic field density in the tape is around 260 mT. As it is shown in Fig. \ref {Fig. 5} for $I_{c}(B,\theta)$ data under different magnetic field values for the modeled tape, this amount of magnetic field causes significant reduction of critical current density. Fig. \ref {Fig. 6} depicts the comparison of the current density in the case of constant $J_c$ (red solid line) and dependent $J_{c}(B,\theta)$ (blue dashed line) when the magnet is just on the top of the tape with the airgap of $3.3$ mm. In the case of constant $J_{c}$, the critical current density remains the same in the width of the tape (green dashed line in Fig. \ref {Fig. 6}), so the overcritical eddy currents that results in generation of electric field and thus voltage belong to the regions where the red solid line exceeds the green dashed lines. However, in the case of considering $J_{c}(B,\theta)$ dependence, the value of $J_{c}$ will diminish up to around 0.8 of $J_{c}$ in the middle part (yellow dashed line in Fig. \ref {Fig. 6}), due to perpendicular magnetic field density up to 260 mT. Therefore, the overcritical current occurs in regions where the blue dashed line exceeds the yellow dashed lines. Experiments and modeling resluts confirm that the latter leads to higher amount of electric field and voltage generated by the flux pump.   

\section{Airgap Dependence of Open-Circuit Voltage}

\begin{figure}[tbp]
\centering
{\includegraphics[trim=0 0 0 0,clip,width=8 cm]{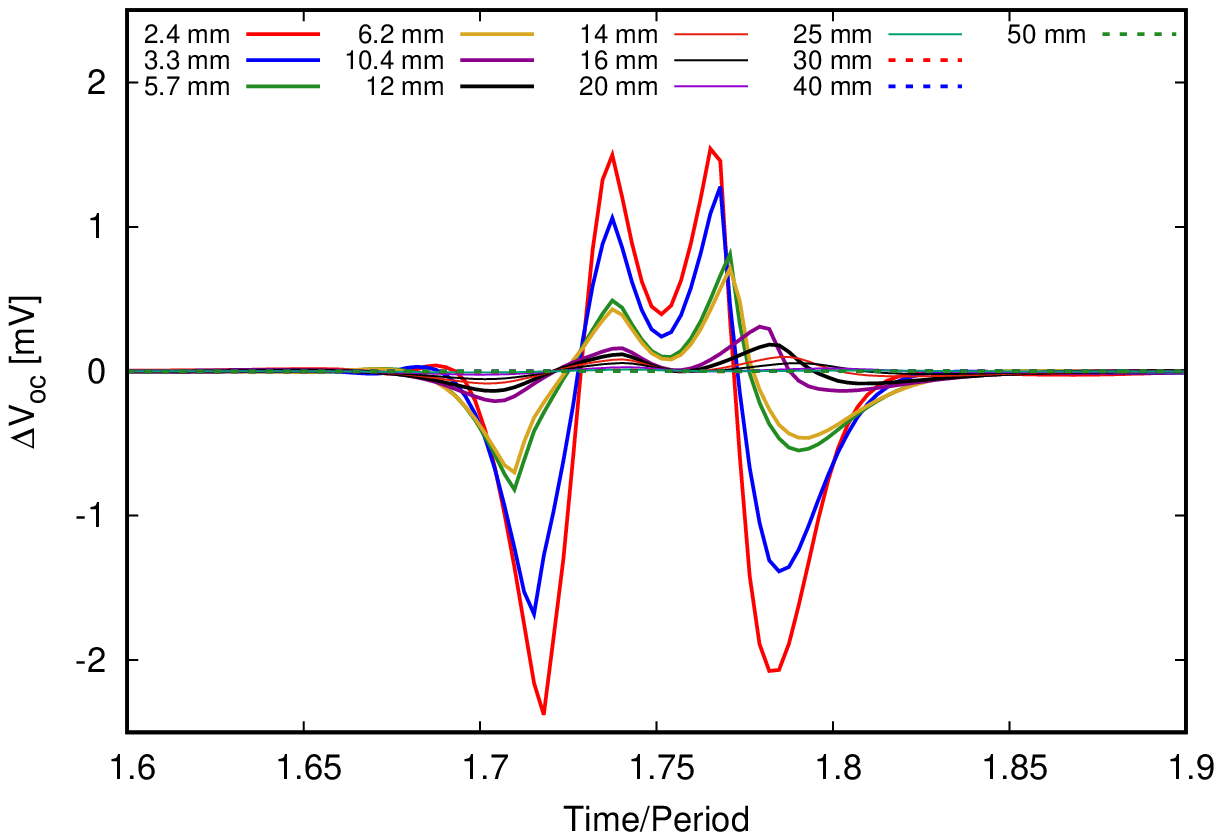}}
\caption {Trend of reduction of $\Delta V_{oc}$ with increasing airgap up to $50$ mm ($\Delta V_{oc} \approx l[\partial_t A_{av,J}+E_{av}(J)]$).} 
\label{Fig. 71}
\end{figure}

 The dc component of open-circuit voltage is the same as that of $ \Delta V $ as defined in Equation (\ref{DeltaV}), since the normal-state contribution at room temperature vanishes. This is convenient because measurements can easily determine $ \Delta V $, which in open-circuit mode can also calculated from Equation (\ref{deltav}).   
 As it is illustrated and showed by experiments in \cite{bumby2016anomalous} and \cite{jiang2015impact}, $\Delta V_{oc}$ decreases with increasing airgap, which is the result of reduction of magnetic field density in the tape. Fig. \ref {Fig. 71} shows the same trend of reduction of $\Delta V_{oc}$ for the modeled flux pump from 2.4 mm up to 50 mm.  
The shape of $\Delta V_{oc}$ also qualitatively agrees with the experiments of \cite{mataira2019origin} and in particular, the increase of the second positive peaks compared to the first and the shift of the positive peaks to the right.    

\begin{figure}[tbp]
\centering
{\includegraphics[trim=0 0 0 0,clip,width=8 cm]{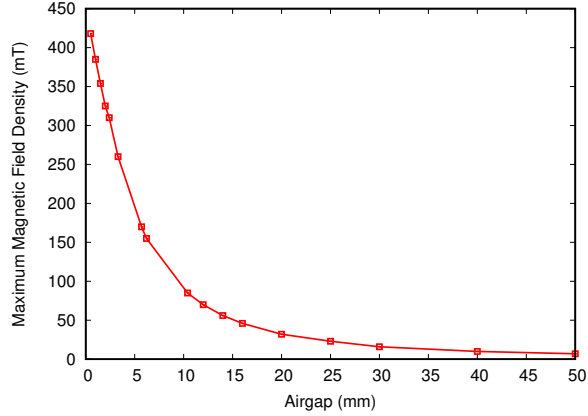}}
\caption {Trend of reduction of maximum perpendicular magnetic field density in the tape with increasing airgap } 
\label{Fig. 7}
\end{figure}

The magnitude of magnetic field density plays an important role in creating voltage in the flux pump. Fig. \ref {Fig. 7} depicts the trend of reduction of maximum perpendicular magnetic flux density in the tape with increasing airgap in the range of 0.5 to 50 mm. As it is obvious, the maximum perpendicular magnetic flux density decreases sharply, so that the reduction of $\Delta V_{oc}$ with increasing airgap is predictable.   

\begin{figure}[tbp]
\centering
{\includegraphics[trim=0 0 0 0,clip,width=8 cm]{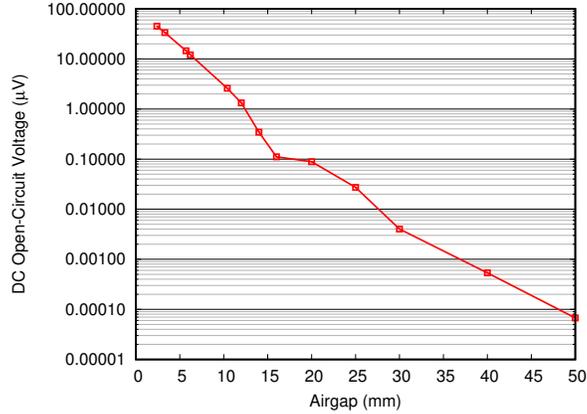}}
\caption {Logarithmic diagram of DC value of open-circuit voltage with the change of airgap for the range of $2.4$ to $50$ mm } 
\label{Fig. 8}
\end{figure}

However, we may face the question that what is the boundary for generating the voltage in a flux pump. In other words, what is the maximum airgap that results in non-zero voltage in the flux pump. This issue is directly related to the value of magnetic flux density on the tape surface. Fig. \ref {Fig. 8} demonstrates the change of open-circuit DC voltage generated in the modeled flux pump for airgaps up to 50 mm.     

\begin{figure}[tbp]
\centering
{\includegraphics[trim=0 0 0 0,clip,width=8 cm]{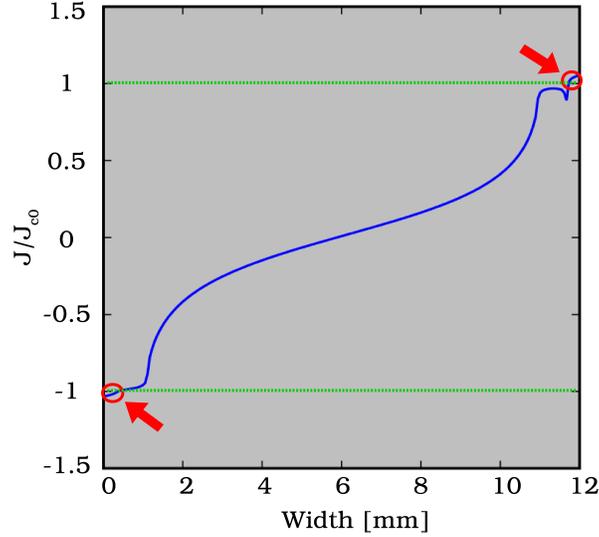}}
\caption {The current density line diagram when the magnet is just on the top of the tape with airgap of $50$ mm. The encircled areas show the overcritical regions of current density and the green dashed lines show the critical current density along the tape width} 
\label{Fig. 9}
\end{figure}

Fig. \ref {Fig. 8} indicates that even in airgap as large as 50 mm, DC voltage still exists considering the maximum perpendicular magnetic field density of 7 mT in the tape. Although this value is insignificant, \textit{i.e.} around 0.0002 $\mu$V, it suggests that there is still voltage generated in the flux pump. Considering this fact that the maximum perpendicular magnetic field density of 7 mT is much less than the penetration field ($B_{p}$) estimated for the modeled HTS tape in the flux pump (\textit{i.e.} 25 mT) \cite{hoffmann2010flux}, it can be concluded that the generation of voltage in flux pump does not require full penetration of the field into the tape, but only partial penetration can lead to voltage generation. For confirming this claim, Fig. \ref {Fig. 9} shows the line diagram of current density with airgap of 50 mm, when the magnet is just on top of the tape.

Fig. \ref {Fig. 9} indicates that the full penetration of the magnetic field has not occurred in the tape when the magnet is on top and in the closest distance to the tape. However, there are some small areas where the current density value exceeds the critical current density of the tape. Note that in here, due to the very small perpendicular magnetic field in the tape, the critical current density value is almost the same as the critical current density at self-field $J_{c0}$ (green dashed lines in Fig. \ref {Fig. 9}). These areas are responsible for generating the low value of voltage observed in Fig. \ref {Fig. 8}.   

\section{Quantitative Comparison to Experiments}

\begin{figure}[tbp]
\centering
{\includegraphics[trim=0 0 0 0,clip,width=8 cm]{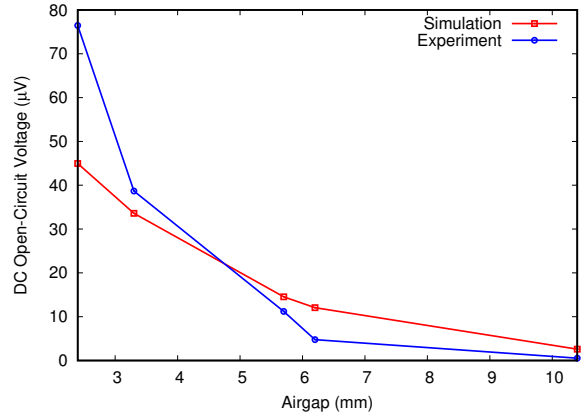}}
\caption {Comparison of modeling results with experimental studies performed in \cite{bumby2016anomalous} for the DC value of open-circuit voltage for different airgaps} 
\label{Fig. 10}
\end{figure}

For confirmation of the modeling results, the DC value of the open-circuit voltage for different airgaps have been compared to experimental studies obtained from \cite{bumby2016anomalous}, which can be observed in Fig. \ref {Fig. 10}.

From Fig. \ref {Fig. 10}, it is apparent that there is good agreement between DC values for airgaps equal to or larger than 3.3 mm. However, there is a big discrepancy in airgap of 2.4 mm. The reason for this discrepancy can be explained by several factors. The first one as explained before in section \ref{section 3.1} comes back to the shape of magnet. The shape of magnet used in experiments was cylindrical while in 2D it is not possible to model this shape and, instead, the square bar shape has been used. This difference becomes bolder when the magnet gets very close to the tape because the shape of flux lines in very close distances are different in cylindrical and infinite rectangular magnets (the modeled one). The other error originates from measurements. This mostly happens due to error in measuring airgaps because of thermal contraction and also changing airgap value during movement of the magnet. In the end, also the modeling error cannot be neglected. This partly originates from the limitations of 2D modeling and assuming infinite tape and magnet in z direction (modeling was performed in xy-plane), so the flux lines are not identical to the real magnet. The variations of $J_{c}(B,\theta)$ whithin the real tape will also cause discrepancy. 

\section{Conclusions}

In this work, a numerical method based on MEMEP has been utilized to model a dynamo-type flux pump. This model is efficient, fast, and provides the opportunity to model more complex geometries and parameter sweeps. Since the principle of the flux pump is not fully understood yet, studying open-circuit mode, which is a simpler case compared to full-circuit mode, can help to better understand its mechanism. For verification of the modeling results under open-circuit condition, they have been compared with experimental studies obtained from \cite{bumby2016anomalous}. With studying the DC value of the open-circuit voltage in the range between 0.5 to 50 mm, it was concluded that the DC voltage generation does not cease even in large airgaps up to 50 mm, where the maximum perpendicular magnetic field density is around 7 mT. This occurs because even under small values of magnetic field, there are still some regions near the edges that current density exceeds the critical current density and leads to the generation of DC voltage in the HTS tape. Furthermore, the comparison of modeling results in the airgap range between 2.4 mm to 10.4 mm showed good agreement with experiments. 

\section*{References}
\bibliographystyle{unsrt}


\end{document}